\documentclass[conference,letterpaper]{IEEEtran}

\usepackage[utf8]{inputenc} 
\usepackage[T1]{fontenc}
\usepackage{url}
\usepackage{ifthen}
\usepackage{cite}
\usepackage[cmex10]{amsmath} 
\usepackage{amssymb}

\usepackage{graphicx}

\newtheorem{theorem}{Theorem}
\newtheorem{definition}{Definition}
\newtheorem{lemma}{Lemma}
\newtheorem{example}{Example}

\begin{document}
\title{Directed Redundancy in Time Series} 



\author{%
  \IEEEauthorblockN{Jan {\O}stergaard}
  \IEEEauthorblockA{Department of Electronic Systems\\
                    Aalborg University\\
                    Aalborg, Denmark\\
                    Email: jo@es.aau.dk}
}


\maketitle


\begin{abstract}
We quantify the average amount of redundant information that is transferred from a subset of relevant random source processes to a target process. 
To identify the relevant source processes, we consider those that are connected to the target process and in addition share a certain proportion of the total  information causally provided to the target. Even if the relevant processes have no directed information exchange between them, they can still causally provide redundant information to the target. This makes it difficult to identify the relevant processes. To solve this issue, we propose the existence of a hidden redundancy process that governs the shared information among the relevant processes. We  bound the redundancy by the minimal average directed redundancy from the relevant processes to the target, from the hidden redundancy process to the target, and from the hidden redundancy process  to the relevant processes. 
\end{abstract}

\section{INTRODUCTION}
\IEEEPARstart{M}{utual information} 
 is frequently used when quantifying the amount of redundancy between random variables \cite{Cover:2006}. 
 Shannon's mutual information measure is only able to quantify the shared information between \emph{two} sets of variables \cite{yeung:2002}. Thus, a naive way to compute the amount of common information of three or more variables is to use the minimum of their \emph{pairwise} mutual informations. Alternative ways of analyzing the intrinsic relationships between multiple random variables have been considered in e.g.,
 \cite{Williams:2010,Harder:2013a,Bertschinger:2014,Ince:2017,James:2017,wibral:2017,Kunert-Graf:2020,james:2021,mediano:2021}. 
 

Consider the case where $Z$ is the target signal and $X$ and $Y$ the source signals. We are interested in assessing the amount of redundant information in $Z$ due to $X$ and $Y$, i.e., the amount of shared information that both $X$ and $Y$ convey to $Z$. This  was considered for discrete random variables in \cite{Williams:2010} as:
\begin{align}\label{eq:Imin}
I^{\mathrm{min}}(X;Y;Z) &\triangleq\! \!\sum_{z\in \mathcal{Z}} p_Z(z) \min ( I(X; Z=z) , I(Y;Z=z)), \\ \notag
I(X;Z=z) &= \sum_{x \in \mathcal{X}} p_{X|Z}(x | z) \log_2 \frac{p_{X|Z}(x|z)}{p_Z(z)}, \\ \notag
I(Y;Z=z) &= \sum_{y \in \mathcal{Y}} p_{Y|Z}(y | z) \log_2 \frac{p_{Y|Z}(y|z)}{p_Z(z)},
\end{align}
where $\mathcal{X},\mathcal{Y},\mathcal{Z}$ are the alphabets of $X,Y$, and $Z$, respectively.
%
%
%

An extension of the redundancy measure in \eqref{eq:Imin} to continuous-alphabet sources was considered in \cite{Barrett:2015}:
\begin{equation}\label{eq:cont}
I^{\mathrm{min}}(X;Y;Z) \!\triangleq \!\!\int_{z \in \mathcal{Z}} \!\!\! \!\!\!\! \! \!\!f_{Z}(z)  \min ( I(X; Z=z) , I(Y;Z=z)) \, dz. 
\end{equation}
%
In order to extend \eqref{eq:cont} to processes, one could use a directed measure for discrete-time stationary processes $\{X_k\}, \{Y_k\},$ and $\{Z_k\}$; such as  transfer entropy $\mathbb{TE}$ \cite{schreiber:2000}:
\begin{align}\label{eq:TE}
\mathbb{TE}(\{X_k\}\to \{Z_k\}) \triangleq
I(X^{n-1} ; Z_n | Z^{n-1} ). 
\end{align}
We can then form a directed version of the average minimal redundancy measure for processes as follows:
\begin{align}\label{eq:red_te}
&I^{\mathrm{min}}(\{X_k\};\{Y_k\}; \{Z_k\}) \triangleq 
\int_{z^{n}\in \mathcal{Z}^n} f_Z(z_n|z^{n-1}) \times \\ \notag
&
\min (  \mathbb{TE}(\{X_k\}\!\to\! \{Z_k=z_k\} , \mathbb{TE}(\{Y_k\}\!\to\! \{Z_k=z_k\})
\, dz^n.
\end{align}
We are, however, facing a number of problems with \eqref{eq:red_te}:
\begin{enumerate} 
\item[\emph{i)}] Computing $\mathbb{TE}(\{Y_k\}\to \{Z_k=z_k\})$ for every $z^k$ is practically infeasible since $\mathcal{Z}^n$ is uncountably infinite. 
\item[\emph{ii)}] If there are more than two source processes, the minimum would be taken over all sources processes. However, perhaps only a few of them are connected to the target, and hence the minimum transfer entropy over \emph{all} source processes and to the target is then always zero.  
\item[\emph{iii)}] A positive transfer entropy does not necessarily imply that the processes are causally related.  
\end{enumerate}
Note also that two source processes that have zero causal interaction with each other, could potentially still causally provide redundant information to the target.

It was suggested in \cite{Barrett:2015} and further developed in \cite{faes:2017} to use the minimum transfer entropy from the sources to the target as a measure of redundancy. Such an approach would solve problem \emph{i)}.
Moreover, the idea that source variables can be weakly or even uncorrelated and yet carry redundant information was discussed in \cite{Bertschinger:2014} and \cite{Barrett:2015}. Here they suggested to not consider the mutual information among all the source variables but only between the sources and the target. We will leverage upon both of these ideas in the sequel. 



In this paper, we propose and analyze a new notion of \emph{directed} redundancy that is applicable to both discrete-time \emph{continuous-}alphabet and \emph{discrete}-alphabet \emph{processes}. The notion is motivated by the analysis of brain networks, where it is currently unknown to what degree the human brain relies upon redundant information for processing \cite{luppi:2022}.

Our main contribution is to quantify the redundant information exchange a set of \emph{relevant} source processes each provide to a given target process. This is complicated by the following problem. Two causally independent sources could provide different (and hence non-redundant) information to the target. Thus, if we only measure the \emph{amount} of causal information exchange from the sources and to the target, then these sources would wrongly be classified as providing redundant information transfer to the target. On the other hand, the 
causal information exchange between the source processes could be zero even though they provide redundant information to the target. Thus, it is necessary but non-trivial to include the dependency between the sources to guarantee that they actually provide redundant information to the target. 
To tackle this problem, 
 we hypothesize the existence of a (perhaps hidden) redundancy process, $\phi$, which causally affect the source processes. Thus, if we know $\phi$, we can compute the causal information exchange from $\phi$ to all the relevant sources, and also to the target. The minimum of these causal exchanges, would be an upper bound to the redundancy. 

\subsection{Notation}
We use $\{X_k\}$ to denote a one-sided scalar random process, i.e., $\{X_k\} = X_1, X_2, \dotsc,$ where $X_k \in \mathbb{R}$. For brevity, we will occasionally refer to $\{X_k\}$ simply by $X$. For example, $\mathbb{TE}(X\to Z)$ denotes the transfer entropy from $\{X_k\}$ to $\{Z_k\}$.
The $i$th element of $\{X_k\}$ is denoted $X_i$, and $X^N = X_1,\dotsc, X_N$. 
We use the notation $X^N(j), j=1,\dots, J,$ to distinguish between $J$ time-series each of length $N$. 
If $X_{k+1} = f_k(X^k,Y^k), \forall k$, for a deterministic function $f_k$, then there is a directional (or causal) dependency from $Y$ to $X$. 

\section{Redundancy via Minimal Sufficient Statistics}
In this section, we motivate the use of minimal sufficient statistics as a means to bound the common information. 
Let  $A,B$, and $C$ be mutually independent discrete random variables. Moreover, let $X=(A,B), Y=(A,C)$, and $Z=(B,C)$. 
What is the amount of common (redundant) information between $X,Y$, and $Z$? 
Using the minimum of their pairwise mutual informations yields:
\begin{align}\label{eq:naive}
\min (I(X;Y), I(X;Z), I(Y;Z))\! =\! \min (H(A), H(B), H(C)),
\end{align}
since $I(X;Y) = H(X) - H(X|Y) = H(A,B) - H(A,B|A,C) = H(A)+H(B) - H(B|A,C) = H(A)$. Similarly, $I(X;Z) = H(B)$ and $I(Y;Z)=H(C)$. Thus, if the entropies are positive the pairwise common information is positively bounded away from zero. 
On the other hand, we can define a notion of redundancy based upon sufficient statistics, which is zero for the above example.
Let $T_{XY}$ be a sufficient statistics of $X$ with respect to $Y$, i.e.,
$I(X;Y) = I(T_{XY};Y)$ and $X - T_{XY} - Y$ forms a Markov chain, i.e., $X$ and $Y$ are conditionally independent of each other given knowledge of $T_{XY}$ \cite{Cover:2006}.
A minimal sufficient statistics $T^*_{XY}$ satisfies \cite{Cover:2006}: 
\begin{equation*}
    T^*_{XY} = \arg\min_{T_{XY}} I(T_{XY} ; X), \quad s.t.\quad  I(T_{XY} ; Y) = I(X;Y).
\end{equation*}

For the above example, one may observe that the minimal sufficient statistics are $T_{XY}=A, T_{XZ}=B$, and $T_{YZ}=C$. Then, clearly $I(T_{XY};T_{XZ}) = I(T_{XZ};T_{YZ}) = I(T_{XY};T_{YZ}) = 0$. Since the minimal sufficient statistics have no information overlap with each other one could argue that the triplet $(X,Y,Z)$ therefore should have zero information in common, which implies zero redundancy. The above example inspire us to introduce a new notion of redundancy for random variables, which we extend to processes in the next section:
\begin{align}\label{eq:bound}
&\bar{I}^{\mathrm{red}}(X;Y;Z) \\ \notag
&\quad \triangleq \min\, ( I(T_{XY};T_{XZ}), I(T_{XZ};T_{YZ}), I(T_{XY};T_{YZ}) ),
\end{align}
where $T_{XY}$ denotes a minimal sufficient statistics for $X$ with respect to $Y$, and similarly for $T_{XZ}$ and $T_{YZ}$. Clearly, $\bar{I}^{\mathrm{red}}(X;Y;Z) \leq \min\, (I(X;Y), I(X;Z), I(Y;Z)).$



\section{Directed Redundancy}
In this section, we   introduce a hidden redundancy process, and  define a notion of directed redundancy for processes. 

\subsection{Hidden redundancy process}\label{sec:example}
The following example illustrates the usefulness of considering a (perhaps hidden) redundancy process that governs the shared information, which is exhanged by a set of source processes to a target process. 
Consider the following system: 
{\allowdisplaybreaks\begin{align}\label{eq:dyn1}
\psi_{k+1} &= a \psi_{k} + W^\psi_k \\
\phi_{k+1} &= a \phi_{k} + W^\phi_k \\
X_{k+1} &= b X_{k} + c\phi_{k-1} + W^X_k \\
Y_{k+1} &= b Y_{k} + c\phi_{k-1} + W^Y_k \\ \label{eq:dyn5}
Z_{k+1} &= b Z_{k} + d X_{k-1} + d Y_{k-1} + e\psi_{k-1} + W^Z_k, 
\end{align}
where $W_k^{\varphi}\in \mathbb{R}, \forall k, \forall \varphi\in \{\phi,\psi,X,Y,Z\}$, are mutually independent and standard normally distributed. }
This set of processes is constructed such that $\phi$ represents a (hidden) redundancy process which "induces" redundant information exchanges to $Z$ via $X$ and $Y$ but not via $\psi$. However, $\psi$ is also causally providing information to the target. Thus, if one would use a measure like \eqref{eq:naive}, then the redundancy would be estimated as being zero, since $\psi$ is causally independent of $X$ and $Y$. Even if the mutual dependency between the sources were left out, the estimate would be the minimum of $I(\psi; Z),I(\phi; Z),I(X; Z),I(Y; Z)$, which is  also wrong, since the information provided by $(X,Y)$ is independent (and hence not redundant) of that provided by $\psi$ to the target. 
Our key idea is to only focus upon the relevant source processes, $X$ and $Y$, which provide some common directed information to $Z$ even though they are causally independent of each other. 

\subsection{Directed Redundancy for Processes}
In the sequel, we will consider the transfer entropy to be our desired measure of directed (causal) information exchange. 
Our results are also valid if the transfer entropy is substituted by, e.g.,  conditional directed information~\cite{massey:1990,Kramer:2003}.


\begin{definition}\label{def:red}
\emph{The transfer-entropy measurable causal redundancy provided to the target $\{Z_k\}$ via the source processes $\{X_k\}, \{Y_k\}$, which are driven by the hidden redundancy process $\{\phi_k\}$  is defined as:}
\begin{align*}
    &I^{\mathrm{red}}(\{X_k\};\{Y_k\}; \{Z_k\}) \triangleq \\
    &\min \ (
      \mathbb{TE}(\phi\to Z), \mathbb{TE}(\phi\to T_{XZ}),\mathbb{TE}(\phi\to T_{YZ}), \\
      &
\mathbb{TE}(T_{XZ} \to Z), \mathbb{TE}(T_{YZ} \to Z),  
    I( \{T_{XZ}\} ; \{T_{YZ}\}) 
    \ ), 
\end{align*}
\emph{where $T_{XZ}$ is a  \emph{causal} minimal sufficient statistics for $X$ with respect to $Z$, which we define to mean that $T_{XZ}$ satisfies the conditional Markov chain}
\begin{equation*}
\vspace{-3mm}
X^{n-1}|_{Z^{n-1}} - T_{XZ}^{n-1}|_{Z^{n-1}} - Z_n|_{Z^{n-1}}, \forall n,
\end{equation*}
where
\begin{align*}
T^{n-1}_{XZ} &= \arg\min_{S_{XZ}^{n-1}} I(S_{XZ}^{n-1}; Z_n | Z^{n-1}) \\
& s.t.\quad I(S_{XZ}^{n-1}; Z_n | Z^{n-1}) = I(X^{n-1} ; Z_n | Z^{n-1}), \forall n. 
\end{align*}
\end{definition}
The last quantity $I( \{T_{XZ}\} ; \{T_{YZ}\})$ in the definition is a standard mutual information. This term is necessary to exclude the case where $X$ and $Y$ would only be providing non-redundant information to $Z$ in which case $I( \{T_{XZ}\} ; \{T_{YZ}\})=0$.

It is   non-trivial to find minimal sufficient statistics for general sources, and below we introduce an upper bound to the redundancy that does not rely on minimal sufficient statistics. 
\begin{lemma}\label{lem:ub1}
\emph{Consider the dynamical system in \eqref{eq:dyn1} -- \eqref{eq:dyn5}. The amount of redundant transfer entropy communicated via $X$ and $Y$ to $Z$, and contained in $Z$, is upper bounded by:}
\begin{align}\notag
I^{\mathrm{red}}(\{X_k\}&;\{Y_k\}; \{Z_k\})  \leq \min\, ( \mathbb{TE}(\phi\to Z), \mathbb{TE}(\phi\to X), \\ \label{eq:lem1}
&\mathbb{TE}(\phi\to Y), \mathbb{TE}(X\to Z),\mathbb{TE}(Y\to Z)).
\end{align}
\end{lemma}
\begin{IEEEproof}
The proof follows by replacing the minimal sufficient statistics in Definition \ref{def:red} by their associated random processes. Omitting $I( \{T_{XZ}\} ; \{T_{YZ}\})$ can only
increase the estimated redundancy. 
 $\mathbb{TE}(X\to Z)$ may overestimate the amount of information exchanged from $X$ to $Z$, since $Z$ also receives information from $Y$. Similarly for $\mathbb{TE}(Y\to Z)$. Thus, it is necessary to also take into account the amount of redundant information that is induced into $X$ and $Y$. This is upper bounded by $\mathbb{TE}(\phi\to X)$ and  $\mathbb{TE}(\phi\to Y)$, respectively. Finally, since $\phi$ is driving the redundancy in the system, the redundancy can be no greater than $\mathbb{TE}(\phi\to Z)$. 
\end{IEEEproof}

The following lemma demonstrates the existence of systems that are bounded by the different terms in \eqref{eq:lem1}, and  thereby showing that  all of the terms are needed.
\begin{lemma}\label{lem:ub2}
    \emph{Consider the system in \eqref{eq:dyn1} -- \eqref{eq:dyn5} and let $a=b=0$, then:}
    \begin{align}\label{eq:phi2x}
      \mathbb{TE}(\phi\to X) &= \mathbb{TE}(\phi\to Y) =  \frac{1}{2}\log_2\big( c^2\sigma_\phi^2 + 1\big), \\ \label{eq:phi2z}
     \mathbb{TE}(\phi\to Z) &=  \frac{1}{2}\log_2\bigg( \frac{4d^2c^2\sigma_{\phi}^2 + 2d^2 + e^2 +1}{2d^2 + e^2 +1} \bigg), \\ \notag
      \mathbb{TE}(X\to Z) &=\mathbb{TE}(Y\to Z) \\ \label{eq:x2z}
      &= \frac{1}{2}\log_2\bigg( \frac{4c^2d^2\sigma_\phi^2 + 2d^2 + e^2 +1}{ \frac{dc^2\sigma_\phi^2}{c^2\sigma_\phi^2+1}+ d^2 +e^2 +1}\bigg),
    \end{align}
    \emph{where $\sigma_\phi^2$ is the stationary variance of $\phi$, and $\sigma^2_{W^{\varphi}}=1$ for $\varphi \in \{\psi,X,Y,Z\}$.
    Moreover, for $c=d>0$, and $e=1$, the minimum in \eqref{eq:lem1} is explicitly given by: }
    \begin{align}\label{eq:cases}
     &I^{\mathrm{red}}(\{X_k\};\{Y_k\}; \{Z_k\})  \\  \notag
     &\leq\begin{cases}
      \mathbb{TE}(\phi\to Z), &   0 < c< 1, \sigma_\phi^2 <4, \\ \notag
      \mathbb{TE}(\phi\to X) = \mathbb{TE}(\phi\to Y),&  c\geq 1, \sigma_\phi^2 <4, \\ \notag
      \mathbb{TE}(X\to Z) = \mathbb{TE}(Y\to Z), &  \xi_1\leq c\leq \xi_2 , \sigma_\phi^2 \geq 4,
    \end{cases}
    \end{align}
    \emph{where $\xi_1 = \frac{1}{2}\sigma_\phi - \frac{1}{2}\sqrt{\sigma_\phi^2 - 4}$ and $\xi_2 = \frac{1}{2}\sigma_\phi + \frac{1}{2}\sqrt{\sigma_\phi^2 - 4}$. }
\end{lemma}
\begin{IEEEproof}
We first expand $\mathbb{TE}(\phi\to Z)$ by the chain rule of mutual information:
    \begin{align*}
        \mathbb{TE}(\phi\to Z)=&I(\phi^{k-1}; Z_k | Z^{k-1}) =  I(\phi_{k-4}; Z_k | Z^{k-1}) \\
        &+ I(\phi^{k-1}\backslash \phi_{k-4}; Z_k | Z^{k-1}, \phi_{k-4}).
        \end{align*}
The first term can be further written as:
{\allowdisplaybreaks
        \begin{align*}
        &I(\phi_{k-4}; Z_k | Z^{k-1}) \\
        &= I(\phi_{k-4};  d(X_{k-2}+Y_{k-2}) + e \psi_{k-2} + W_{k-1}^Z | Z^{k-1}) \\
        &=  I(\phi_{k-4};  d( 2c\phi_{k-4} + W_{k-3}^{X} + W_{k-3}^{Y} ) + e \psi_{k-2} \\
        &\quad + W_{k-1}^Z | Z^{k-1}) \\
        &\overset{(*)}{=} I(\phi_{k-4};  d( 2c\phi_{k-4} + W_{k-3}^{X} + W_{k-3}^{Y} ) + e \psi_{k-2} + W_{k-1}^Z ) \\
        &= \frac{1}{2}\log_2\bigg( \frac{4d^2c^2\sigma_{\phi}^2 + 2d^2 + e^2 +1}{2d^2 + e^2 +1} \bigg).
    \end{align*}
Since $a=b=0$, $(*)$ follows, and it further implies  $I(\phi^{k-1}\backslash \phi_{k-4}; Z_k | Z^{k-1}, \phi_{k-4}) = 0$.} 
    Similarly, we have that:
    \begin{align*}
    &\mathbb{TE}(\phi\to X) = I(\phi^{k-1}; X_k | X^{k-1}) \\
&= I(\phi_{k-2}; c\phi_{k-2} + W_{k-1}^X | X^{k-1}) \\
&\quad + I(\phi^{k-1}\backslash \phi_{k-2}; X_k | X^{k-1}, \phi_{k-2}).\\
        &=I(\phi_{k-2}; c\phi_{k-2} + W_{k-1}^X | X^{k-1}) 
        =  \frac{1}{2}\log_2\big( c^2\sigma_\phi^2 + 1\big) \\
        &=  \mathbb{TE}(\phi\to Y).
    \end{align*}

    Finally,  $\mathbb{TE}(Y\to Z) = \mathbb{TE}(X\to Z)$, where: 
     \begin{align*} \notag
        &\mathbb{TE}(X\to Z)=I( X^{k-1} ; Z_k | Z^{k-1}) \\ \notag
        &= I(X_{k-2} ; d X_{k-2} + d Y_{k-2} + e \psi_{k-2} + W_{k-1}^{Z} | Z^{k-1}) \\ 
        &= \frac{1}{2}\log_2\bigg( \frac{4c^2d^2\sigma_\phi^2 + 2d^2 + e^2 +1}{d\mathrm{var}(c\phi_{k-4} | X_{k-2}) + d^2 +e^2 +1}\bigg), \\
        &\mathrm{var}(c\phi_{k-4}| X_{k-2}) = \frac{c^2\sigma_\phi^2}{c^2\sigma_\phi^2+1}.
    \end{align*}
    All the arguments of the logarithms in \eqref{eq:phi2x}, \eqref{eq:phi2z}, and \eqref{eq:x2z} are convex in $c=d$, and are having their minima at $c=0$ and maxima at $c=\infty$. Pairwise equating the arguments of \eqref{eq:phi2x}, \eqref{eq:phi2z}, and \eqref{eq:x2z}, and solving for $c$, yields  real roots corresponding to the cases provided in \eqref{eq:cases}. The minima can then readily be verified by evaluating the arguments for any $c$ within the corresponding regions. 
\end{IEEEproof}

\begin{example}
Let  $a=d=e=1/3, b=1/5,c=1/2$, and let the length of the processes in \eqref{eq:dyn1} -- \eqref{eq:dyn5} be $N=5000$. The  
transfer entropies between all pairs of processes are shown in Table~\ref{tab:TE}.\footnote{The TRENTOOL transfer entropy estimator  of \cite{lindner:2011} was used.}
It can  be observed that all sources ($\psi, \phi, X,Y$) appears to be coupled to the target $Z$, and  one could therefore wrongly conclude that they all provide a degree of common (redundant) information, which would be upper bounded by $\mathbb{TE}(\psi \to Z) \approx 0.03$. This is clearly a false assessment, since the information provided by $\psi$ to $Z$ is distinct from that provided by $X$ to $Z$ and hence $\psi$ and $X$ convey zero common information to $Z$.
%
Let us now ignore $\psi$ and treat $\phi$ as the source that is driving the redundancy. 
An upper bound to the amount of common information which $X$ and $Y$ both relay to $Z$ is then via Lemma~\ref{lem:ub1} estimated as $\mathbb{TE}(\phi\to Z)\approx 0.04$. 
\end{example}

\begin{table}[ht]
\begin{center}
\caption{Transfer entropies between the processes in \eqref{eq:dyn1} -- \eqref{eq:dyn5}.}
\label{tab:TE}
\begin{tabular}{|c|ccccc|}\hline
From/To & $\psi$ & $\phi$ & $X$ & $Y$ & $Z$    \\ \hline 
$\psi$ & -- &  0.00 &  0.00 &  0.00   & 0.03  \\
$\phi$ & 0.00    &     --  &  0.09  &  0.09   & \textbf{0.04} \\
 $X$ &   0.00  & 0.00   &      --   & 0.00  &  0.06 \\
$Y$ &   0.00 &  0.00   & 0.00     &    --  &  0.06 \\
 $Z$ & 0.00  & 0.00  &  0.00   & 0.00    &     -- \\ \hline
\end{tabular}
\end{center}
\end{table}

In the above example, we deliberately made the choice of omitting $\psi$, and focusing on the exchange of redundant information from $\phi$ only via the sources $X$ and $Y$ to the target $Y$. 
For larger systems, or for systems with unknown dynamics, it would be advantageous to use a systematic approach to select the hidden redundancy process and the associated desired set of relevant sources. 
In principle, one could consider all possible subsets of sources and find the one that maximizes the minimal transfer entropy among them. This is a problem with exponential complexity in the number of possible subsets, and equally important, it is known that the amount of redundant information between  variables is decreasing in the number of variables \cite{Williams:2010}. Thus, such a strategy would potentially be biased towards selecting a minimal number of sources. We propose an efficient strategy for finding a hidden redundancy process and selecting its associated relevant sources in Section~\ref{sec:select}.

\section{Selecting relevant source processes}\label{sec:select}
In this section, we consider multiple sources $\mathcal{S}\triangleq \{X^{N}(1), X^{N}(2), \dotsc, X^{N}(J)\},$ and a single target $Z^N$. 
We will first identify the subset $\mathcal{T} \subseteq \mathcal{S}$ of sources that are relevant with respect to the target $Z^N$, i.e., we are interested in determining whether they are "causally" coupled to the target as measured by the strength of the transfer entropy from $X^{N}(j)$ to $Z^N$. 
Then we identify the hidden process, $\phi$, and finally identify the subset $\mathcal{R} \subseteq \mathcal{T}$ of \emph{relevant} sources, i.e., those that are all governed by the hidden process $\phi$ and in addition they are all coupled to the target. 


\subsection{Target-relevant source processes}
We propose to find a tentative set $\mathcal{T}$ of sources, which is a set of sources that are all coupled to the target. We refer to $\mathcal{T}$ as the target-relevant set of sources. 
We choose to define this set as the one with the minimal number of sources that provides at least an $\eta$-fraction of the total sum of transfer entropy from all sources to the target. This is akin to the smallest high-probability sets in \cite[Ch.4.3]{mackay:2003}.
This can formally be stated as the following optimization problem (for $\eta_T \in [0,1]$):
\begin{align} \label{eq:J}
&\mathcal{T}=\arg \min_{\mathcal{J} \subseteq \{1,\dotsc, J\}} |\mathcal{J}| \\ \notag
& s.t.\ 
\sum_{j\in \mathcal{J}} \mathbb{TE}( X^N(j) \to Z^N)  \geq 
\eta_T \sum_{j=1}^{J}\mathbb{TE}( X^N(j) \to Z^N).
\end{align}
The constraint in \eqref{eq:J} provides an efficient way to omit sources that are only weakly coupled to the target.  
The optimization problem can be efficiently computed by first sorting the transfer entropies in descending order, and then including the greatest elements one by one until the threshold is breached. If several elements have equal probability, there could be ambiguous output unless $\eta_T$ is adjusted to either include or exclude all of the equal probability elements, cf.\ \cite{mackay:2003}.

\subsection{Identifying the hidden process}\label{sec:minimal}



We will now find the hidden redundancy process, which is that particular source out of all the sources in $\mathcal{S}$, which dominates the redundancy within a selective group of \emph{relevant} sources.  
For each source $X^N(i)$, where $ i \in \mathcal{S}$, find the minimal subset $\hat{\mathcal{T}}(i)$ of $\mathcal{T}$ that provides a given fraction $(\eta_H \in [0,1])$ of the total amount of transfer entropy:
\begin{align} \label{eq:Ti}
&\hat{\mathcal{T}}(i) = \arg \min_{\mathcal{J} \subseteq \mathcal{T} } |\mathcal{J}|, \quad s.t. \\ \notag
& 
\sum_{j\in \mathcal{J}} \mathbb{TE}( X^N(i) \to X^N(j)  )  \geq 
\eta_H \sum_{j\in \mathcal{T}}\mathbb{TE}( X^N(i) \to X^N(j) ).
 \end{align}

The minimal causal information exchange  $R(i)$ from the $i$th source and to its associated processes in $\hat{\mathcal{T}}(i)$ is:
\begin{align} 
    R(i) = \min_{j\in \hat{\mathcal{T}}(i)} \mathbb{TE}( X^N(i) \to X^N(j) ).
\end{align}

Out of all the $J$ source processes in $\mathcal{S}$, the one that induces the greatest minimal redundancy within its associated set of sources will be chosen as the hidden redundancy process:
\begin{align}
    i^* &= \arg\max_{i\in \mathcal{S}} R(i), \\ \label{eq:phi}
    \phi^N &= X^N(i^*), \\ \label{eq:hatT}
    \mathcal{R} &= \hat{\mathcal{T}}(i^*),
\end{align}
where $\phi^N$ is the hidden redundancy process, $\mathcal{R} \subseteq \mathcal{T}$ is the set of relevant sources associated with $\phi^N$, and $R(i^*)$ is an upper bound on the redundancy induced by $\phi^N$ into the set $\mathcal{R}$.

\subsection{Bounding the minimal redundant information}
Inspired by Lemma~\ref{lem:ub1}, we formulate the following upper bound to the causal redundancy defined by Definition~\ref{def:red}:
\begin{theorem}
\emph{The  minimal redundant information causally exchanged (as measured by the transfer entropy) to the target via a set of relevant sources, and which is due to a single hidden redundancy process is upper bounded by:}
\begin{align}\notag
I^{\mathrm{red}}(X^N(1);&\cdots; X^N(J); Z^N)  \\
&\leq \min\, ( R_{\phi\to Z},  R_{\phi \to \mathcal{R}} , R_{\mathcal{R} \to Z}   ),
\end{align}
\emph{where}
\begin{align} \label{eq:R1}
    &R_{\phi\to Z} = \mathbb{TE}(\phi^N\to Z^N), 
    R_{\phi \to \mathcal{R}} = R(i^*),\\  \label{eq:R2}
    &R_{\mathcal{R} \to Z} = \min_{j\in \mathcal{R}} \mathbb{TE}( X^N(j) \to Z^N ).
\end{align}
\end{theorem}
\begin{IEEEproof}
$R_{\phi\to Z}$ bounds the information exchange from the hidden redundancy process to the target, 
$R_{\phi \to \mathcal{R}}$ bounds the information exchange from the hidden redundancy process and to the relevant sources, and $R_{\mathcal{R} \to Z}$ bounds the information exchange from the relevant sources to the target. 
\end{IEEEproof}

\section{Simulation study}
We consider 76-channel real-world intracranial EEG recordings obtained from a patient undergoing 8 periods of seizures. 64 cortical electrodes (referred to as electrodes 1 -- 64) and 12 in-depth electrodes (referred to as electrodes 65 -- 76) arranged as two 6 nodes depth-strips were implanted on the subject. 
Details about the data can be obtained from \cite{Kramer:2013}\footnote{Data available online: http://math.bu.edu/people/kolaczyk/datasets.html}. It was pointed out in \cite{Kramer:2013} that in particular the depth-strip with electrodes 71 -- 76 and the lower left corner of the cortical grid with electrodes 1 -- 4, 9 -- 11, and 17 exhibited strong synchronous neuronal activity during seizures. 

\begin{figure}
    \centering
  \includegraphics[width=4.5cm]{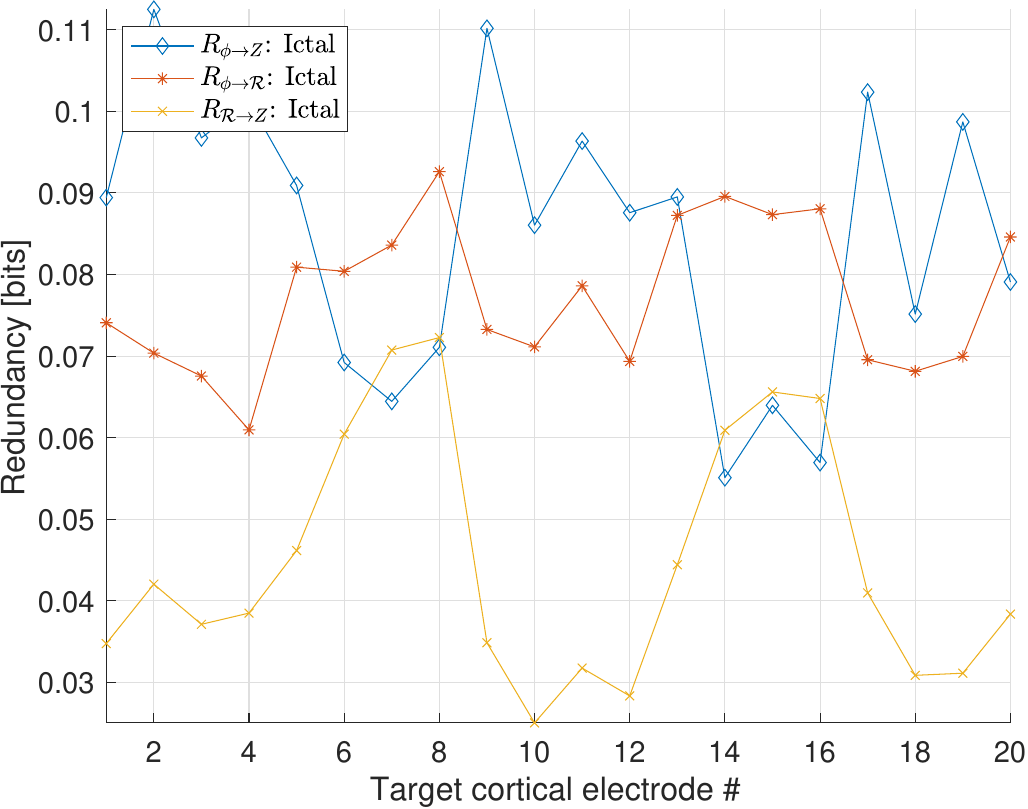}
      \includegraphics[width=4cm]{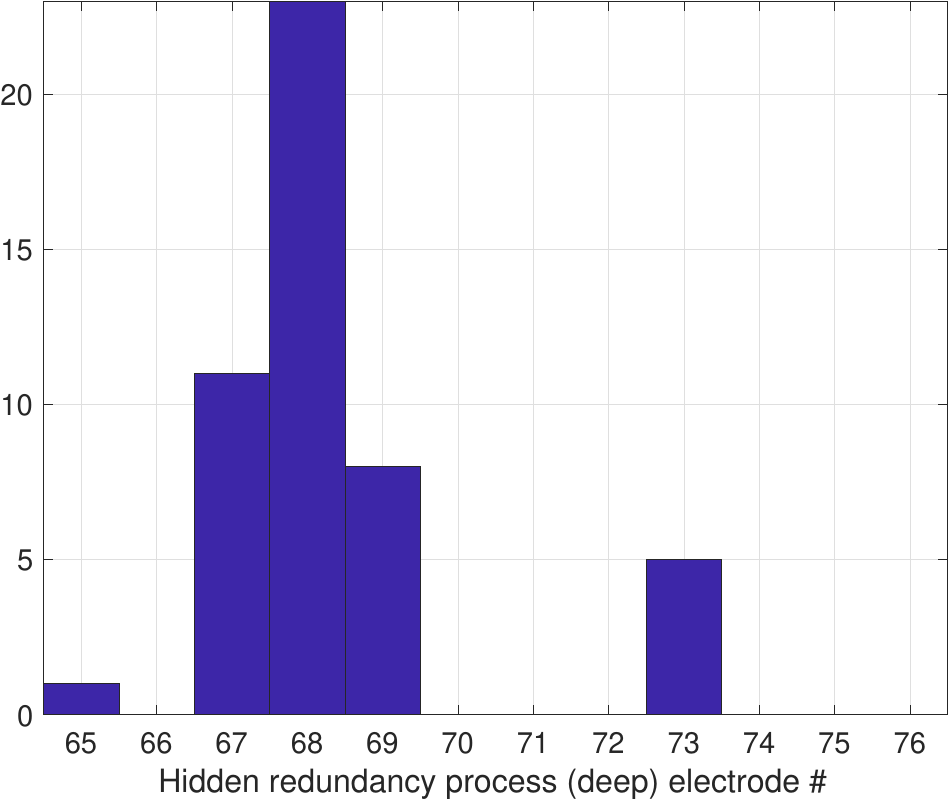}
    \caption{\textbf{Left:} Redundancy \eqref{eq:R1} -- \eqref{eq:R2} in intracranial EEG data. 
    \textbf{Right:} Histogram of the sources that are chosen as the hidden redundancy process for the targets  6 -- 8 and 14 -- 16.}
    \label{fig:redundancy_ecog}\label{fig:hist_redundancy_ecog}
\end{figure}

In our study, the lower range of cortical electrodes ranging from 1 -- 20 will be considered the targets. The sources will be all the 12 depth electrodes and the hidden redundancy processes will also be found amongst the 12 depth electrodes. 
For each target electrode we first find the set of sources $\mathcal{T}$ that are coupled to the target using \eqref{eq:J}. 
Then we identify the hidden redundancy process, which is inducing the majority of the redundancy in the set. This is done using \eqref{eq:phi}. 
%
Having identified a potential hidden redundancy process for a given target electrode, we then use \eqref{eq:hatT} to find the associated set $\mathcal{R}$ of \emph{relevant} source processes. From the target, the hidden redundancy process, and this set of relevant source processes, we can find the corresponding transfer entropies given by \eqref{eq:R1}. These are plotted in Fig.~\ref{fig:redundancy_ecog}. The parameters used for the transfer entropy estimator are: a maximum lag of 5,  10 nearest neighbors, and $\eta_T=\eta_H = 0.8$. A greater $\eta_T$ means less source selected.

\begin{figure}
    \centering
    \includegraphics[width=4.2cm]{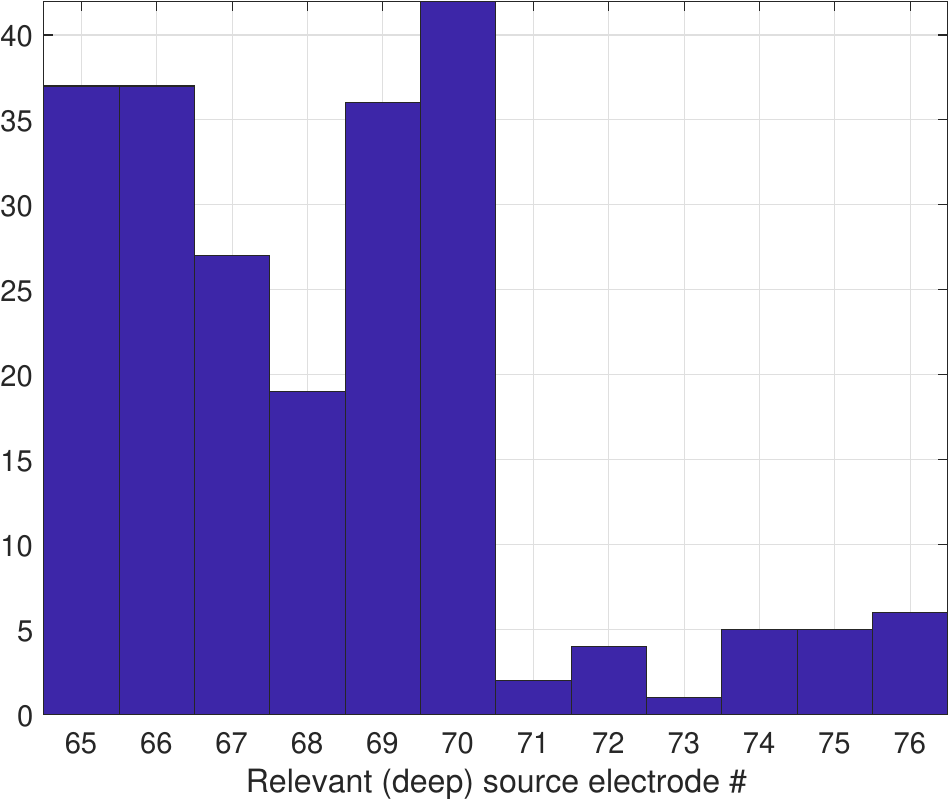} 
      \includegraphics[width=4.2cm]{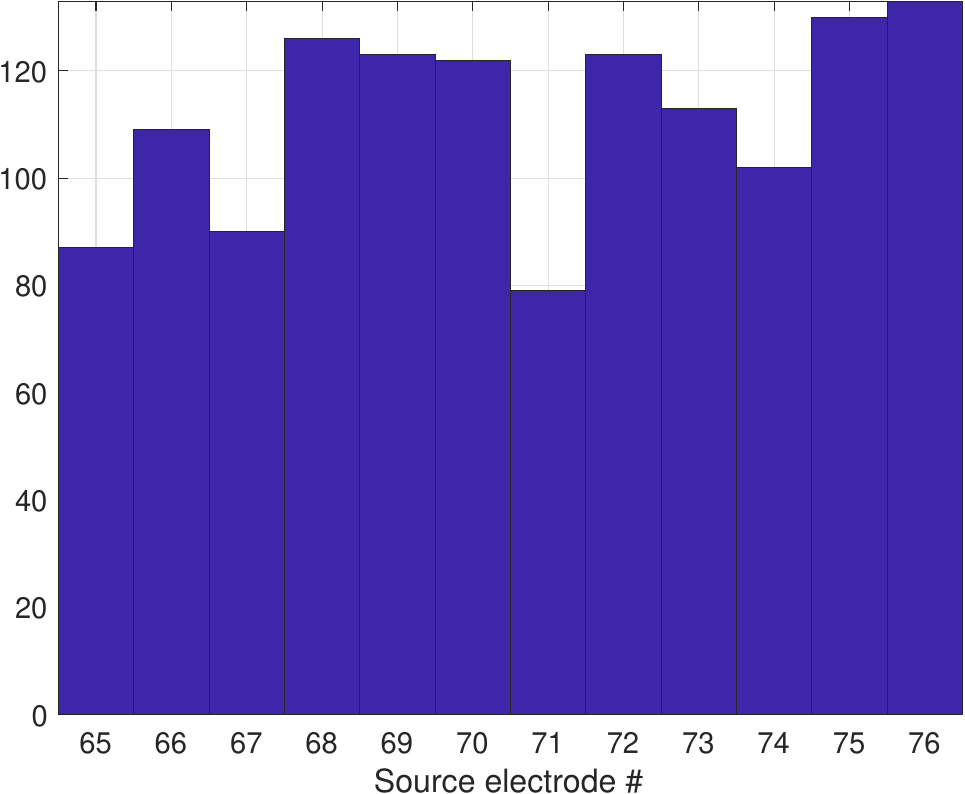}

    \caption{\textbf{Left}: Histogram of the selected relevant sources using a hidden redundancy process.  Only hidden redundancy processes and relevant source processes that are causally coupled to the targets 6 -- 8 and 14 -- 16 are considered. 
        \textbf{Right}: Histogram of $\mathcal{T}$ in \eqref{eq:J}, i.e., the 80\% greatest coupled sources to the targets without using a hidden redundancy process.}
    \label{fig:hist_ecog}
\end{figure}

From Fig.~\ref{fig:redundancy_ecog} one can observe that the minimal redundancy mainly follows $R_{\mathcal{R}\to Z}$, i.e., the  minimal transfer entropy from the relevant sources and to the targets. 
It is interesting to see that the minimal redundancy is peaking for target electrodes 6 -- 8 and 14 -- 16, which are complementary to the electrodes identified in \cite{Kramer:2013} that were exhibiting strong synchronous neuronal activity. Moreover, the transfer entropy $R_{\phi\to Z}$ between the hidden redundancy processes and the targets reaches its minima when $R_{\mathcal{R}\to Z}$ is peaking. This can be explained by the fact that when $R_{\mathcal{R}\to Z}$ is high, the same information is exchanged via different sources, and thus the total information about $\phi$ that exists in the targets is not much greater than that provided by a single source. On the other hand, when $R_{\mathcal{R}\to Z}$ is low, the different sources could potentially provide distinct information about $\phi$ to the target, which would result in an overall greater amount of information about $\phi$ in the targets.  
To further explore the cases, where the redundancy is peaking, i.e., for target electrodes 6 -- 8 and 14 -- 16, we have shown the histograms for the selected hidden redundancy processes, see Fig.~\ref{fig:hist_redundancy_ecog} (right), and their associated relevant sources, see Fig.~\ref{fig:hist_ecog} (left).
It is interesting to note from Fig.~\ref{fig:hist_ecog} (left) that the redundancy appears to mainly come  from within sources on the same depth-strip (electrodes 65 -- 70). Moreover, Fig.~\ref{fig:hist_redundancy_ecog} (right) shows that the hidden redundancy process is also mainly belonging to this depth-strip. The most frequently chosen electrode as the hidden redundancy process is \#68. This one is then selected less as a relevant source, since the hidden redundancy process is excluded when finding the relevant sources. We have tried to compare our findings with the graph theoretic analysis of the electrodes provided in \cite{Kramer:2013}, however, we have been unable to explain why electrode 68 appears to be the main driver of redundancy.   
For comparison, we have in Fig.~\ref{fig:hist_ecog} (right) shown the histogram of the sources $\mathcal{T}$ in \eqref{eq:J}, i.e., those having 80\% of the entire transfer entropy to the targets. In this case, the hidden redundancy process is not taken into account, and a more uniform distribution is obtained compared to Fig.~\ref{fig:hist_ecog}, where the hidden redundancy process is taken into account.

\section{Conclusions}
We proposed a new methodology for quantifying the causal redundancy provided by a set of source processes (time series) to a target process. The sources  do not need to be causally dependent upon each other. Instead we motivated the existence of a hidden redundancy process, which governs the redundancy within a subset of the source processes. We defined a set of relevant source processes to be those that  are causally coupled to the target, and where the hidden redundancy process is causally coupled to all of them. We finally provided an upper bound for the redundancy as the minimal average redundancy from the relevant sources to the target, from the hidden redundancy process to the target, and from the hidden redundancy process  to the relevant sources.


\IEEEtriggeratref{12}



%



\bibliographystyle{IEEEtran}

\bibliography{sample}

\vfill\pagebreak

\end{document}